\documentclass[a4paper,10pt,twoside]{cpc-hepnp}

\usepackage{multicol}
\usepackage{graphicx}
\usepackage{booktabs}
\usepackage{amssymb,bm,mathrsfs,bbm,amscd}
\usepackage[tbtags]{amsmath}
\usepackage{lastpage}
\usepackage{CJK}
\allowdisplaybreaks[4]

\begin{document}
\begin{CJK*}{GBK}{song}

\fancyhead[c]{\small Chinese Physics C~~~Vol. XX, No. X (201X)
XXXXXX} \fancyfoot[C]{\small 010201-\thepage}

\footnotetext[0]{Received 18 November 2014}

\title{Identifying the structure of near-threshold states from the line shape\thanks{This work is supported, in
part, by National Natural Science Foundation of China (Grant Nos.
11147022, 11035006 and 11305137), Chinese Academy of Sciences
(KJCX2-EW-N01), Ministry of Science and Technology of China
(2009CB825200), DFG and the NSFC (No. 11261130311) through funds
provided to the Sino-German CRC 110 ``Symmetries and the Emergence
of Structure in QCD", and Doctor Foundation of Xinjiang University
(No. BS110104) }}

\author{  CHEN Guo-Ying$^{1,4)}$\email{chengy@pku.edu.cn}
\quad HUO Wen-Sheng$^{1)}$\email{uni\_xinjiang@sina.com } \quad ZHAO
Qiang$^{2,3)}$\email{zhaoq@ihep.ac.cn} } \maketitle

\address{%
$^1$ Department of Physics, Xinjiang University, Urumqi 830046,
China \\
$^2$ Institute of High Energy Physics, Chinese Academy of Sciences,
Beijing 100049, China \\
$^3$ Theoretical Physics Center for Science Facilities, CAS, Beijing
100049, China\\
$^4$ State Key Laboratory of Theoretical Physics, Institute of
Theoretical Physics, Chinese Academy of Sciences, Beijing 100190,
China }

\begin{abstract}
We revisit the compositeness theorem proposed by Weinberg in an
effective field theory (EFT) and explore criteria which are
sensitive to the structure of $S$-wave threshold states. On a
general basis, we show that the wave function renormalization
constant $Z$,  which is the probability of finding an elementary
component in the wave function of a threshold state, can be
explicitly introduced in the description of the threshold state. As
an application of this EFT method, we describe the near-threshold
line shape of the $D^{\ast 0}\bar D^0$ invariant mass spectrum in
$B\rightarrow  D^{\ast 0}\bar D^0 K$ and determine a nonvanishing
value of $Z$. It suggests that the $X(3872)$ as a candidate of the
$D^{\ast 0}\bar D^0$ molecule may still contain a small $c\bar{c}$
core. This elementary component, on the one hand, explains its
production in the $B$ meson decay via a short-distance mechanism,
and on the other hand, is correlated with the $D^{\ast 0}\bar D^0$
threshold enhancement observed in the $D^{\ast 0}\bar D^0$ invariant
mass distributions. Meanwhile, we also show that if $Z$ is non-zero,
the near-threshold enhancement of the $D^{\ast 0}\bar D^0$ mass
spectrum in the $B$ decay will be driven by the short-distance
production mechanism.
\end{abstract}



\footnotetext[0]{\hspace*{-3mm}\raisebox{0.3ex}{$\scriptstyle\copyright$}2013
Chinese Physical Society and the Institute of High Energy Physics
of the Chinese Academy of Sciences and the Institute
of Modern Physics of the Chinese Academy of Sciences and IOP Publishing Ltd}%

\begin{multicols}{2}

\section{Introduction}

Recently the observations of many new resonances, namely the
so-called $XYZ$ states, have initiated intensive studies of their
properties in both experiment and theory. An interesting feature
about most of these new resonances is that their masses are
generally close to $S$-wave two-particle thresholds and their
coupling to the corresponding $S$-wave two-particle channel is
important. For example, the famous $X(3872)$ is one of the earliest
observed states correlated to the $D^{\ast 0}\bar D^0$ threshold. We
use the notation $D^{\ast 0}\bar D^0$ to denote both $D^{\ast 0}\bar
D^0$ and $\bar D^{\ast 0}D^0$. The structure of these near-threshold
resonances are still under debate and there are various existing theoretical interpretations, which include proposals for
treating them as either conventional quark-antiquark states or QCD
exotics such as tetraquarks, hybrid states, dynamically generated
states or molecular states. In some scenarios, they are treated as
mixing states of the above mentioned configurations. It is worth
mentioning that the recent experimental signals for the charged
quarkonium states $Z_b(10610)$ and $Z_b(10650)$~\cite{Belle:2011aa}
and their analogues in the charmonium sector
$Z_c(3900)$~\cite{Ablikim:2013mio,Liu:2013dau,Xiao:2013iha} and
$Z_c(4020/4025)$~\cite{Ablikim:2013wzq,Ablikim:2013xfr} appear
to be strongly correlated to the thresholds of either $B^{(*)}$ or
$D^{(*)}$ pairs. Since most of those newly observed states are in
the vicinity of an $S$-wave open threshold, a theoretical method to
distinguish whether such a near-threshold state is an elementary
state of overall color singlet or a composite state consisting of
open channel hadrons as constituents is thus crucial for our
understanding of their nature. Although this issue has been
explored for a long time and by many theorists (e.g. one can refer to the early literature of
Refs.~\cite{Salam,Weinberg1,Weinberg2,Lurie} and recent
review~\cite{Brambilla:2010cs} and references therein), our
knowledge about such non-perturbative phenomena is still far from
complete.

The aim of this present work is to develop an effective field
theory (EFT) approach to identify the structure of the near threshold
states and explore its application to the structures of the
recently discovered $XYZ$ states. In particular, we shall study the
$X(3872)$ which, since its first observation by the Belle collaboration
in $B^{\pm}\rightarrow K^{\pm}\pi^+\pi^-J/\psi$~\cite{Choi:2003ue},
has initiated tremendous interest in both experiment and theory.
Due to the small mass difference between the measured
mass of the $X(3872)$ and the $D^{\ast 0}\bar D^0$ threshold, the $X(3872)$ is the best candidate for an $S$-wave $D^{\ast
0}\bar D^0$
molecule~\cite{Tornqvist:2004qy,Close:2003sg,Wong:2003xk,Braaten:2003he,Voloshin:2003nt,Swanson:2003tb,Swanson:2004pp}.
However, it has also been recognized that the structure of the
$X(3872)$ could be rather profound because its large production
rates  in the $B$-factories and at the Tevatron seem to favor a compact
structure in its wave function rather than a loosely bound molecular
state~\cite{charm1,charm2,charm3}. Taking both the production and
decay properties into account, it seems reasonable to identify the
$X(3872)$ to be a mixing state between the $J^{PC}=1^{++}$ $c\bar c$
component and the $D^{0\ast}\bar D^{0}$
component~\cite{charm1,charm2}. This scenario can also explain why
the $\chi_{c1}^\prime$ $c\bar c$ state around 3950 MeV predicted by
the single-channel theory is missing in experiment~\cite{charm4}. It
is worth noting the recent lattice QCD result that a candidate for
the $X(3872)$ about $11\pm 7$ MeV below the $\bar D^0 D^{\ast 0}$
threshold was identified in a lattice simulation with $m_\pi=266(4)$
MeV~\cite{Prelovsek:2013cra}. It was also shown that the pion mass
dependence of the binding energy can provide important information
on the structure of the
$X(3872)$~\cite{Wang:2013kva,Baru:2013rta,Jansen:2013cba}.

Obviously, more experimental data and theoretical development are
required to clarify the nature of the $X(3872)$. Very recently the
LHCb collaboration found evidence for the decay mode
$X(3872)\rightarrow \psi(2S)\gamma$ in $B^+\rightarrow X(3872)K^+$.
The measured ratio of the branching fraction of $X(3872)\rightarrow
\psi(2S)\gamma$ to that of $X(3872)\rightarrow J/\psi\gamma$ is
$R_{\psi\gamma}=\frac{\mathcal{B}(X(3872)\rightarrow\psi(2S)\gamma)}{\mathcal{B}(X(3872)\rightarrow
J/\psi\gamma)}=2.46\pm0.64\pm0.29$~\cite{Aaij:2014ala}. Such a large
value for $R_{\psi\gamma}$ does not support a pure $D^{\ast 0}\bar
D^0$ molecular interpretation of the $X(3872)$, because
$R_{\psi\gamma} $ is predicted to be rather small for a pure
$D^{\ast 0}\bar D^0$ molecular~\cite{Dong:2009uf}.

Since the pure molecular interpretation of the $X(3872)$ is not
favoured, it is then important to study quantitatively how large the
compact component is in the wave function of the $X(3872)$. This is
the main subject of this study. By analyzing the compositeness
relation proposed by Weinberg in the effective field theory, we will
establish the relation between experimental observable and the wave
function renormalization constant $Z$ such that the hadron structure
information encoded in $Z$ can be probed via the measurement of some
of those sensitive observables. Specifically, we will show that the
line shape of $D^{\ast 0}\bar D^0$ in $B\rightarrow
X(3872)(\rightarrow D^{\ast 0}\bar D^0) K $ could be useful for
shedding  important light on the structure of the $X(3872)$, see
also a recent study in Ref.~\cite{ZhengHan}.

\section{Weinberg's compositeness theorem in EFT}

To proceed, we first give a short review of Weinberg's method to evaluate the
coupling constant between a near-threshold state and its
two-particle channel~\cite{Weinberg1,Weinberg2}. Without
losing generality, a total Hamiltonian $H$ of interest can be split
into a free part $H_0$ and an interaction part $V$ to an open
channel near the threshold:
\begin{equation}
H=H_0+V.
\end{equation}
The eigenstates of the free part $H_0$ include the continuum states
$|\alpha\rangle$ and the possible discrete bare elementary particle
states $|n\rangle$, with
\begin{align}
H_0|\alpha\rangle&=E(\alpha)|\alpha\rangle,\ \ \ \ \langle
\beta|\alpha\rangle=\delta(\beta-\alpha),\nonumber\\
H_0|n\rangle&= E_n|n\rangle,\ \ \ \ \langle \alpha|n\rangle=0,\ \ \
\ \langle m|n\rangle=\delta_{m,n},\label{Seq1}
\end{align}
where the energies are defined relative to the two-particle
threshold throughout this paper. The completeness relation for the
eigenstates of $H_0$ reads
\begin{equation}
1=\sum_{n}|n\rangle\langle n|+\int d\alpha |\alpha\rangle\langle
\alpha|.\label{comple}
\end{equation}
A physical bound state $|d\rangle$  is a normalized eignestate of
the total Hamiltonian $H$, with
\begin{equation}\label{Seq2}
H|d\rangle=-B|d\rangle,\ \ \ \langle d|d\rangle=1 \ ,
\end{equation}
where $B>0$ is the binding energy.  We call $|d\rangle$ a
physical bound state in the sense that it has the two open channel
 particles as constituents in its wavefunction and its mass is
below the two-particle threshold or equally, $B>0$. With the
completeness relation in Eq.~(\ref{comple}) and the normalization of
$|d\rangle$ we can have
\begin{align}
1=Z+\int d\alpha|\langle\alpha|d\rangle|^2,\ \ \ \ \ \
Z\equiv\sum_n|\langle n|d \rangle|^2,\label{comple2}
\end{align}
where $Z$ is the probability of finding an elementary state in the
physical bound state. Hence $Z=0$ indicates that the physical bound
state is purely composite, while $0<Z<1$ indicates that there also
exists an elementary component inside the physical state. The
determination of the value of $Z$ would thus enable us to
distinguish a pure composite state from a mixture of a composite and
elementary configuration.

With the relation $|d\rangle=[H-H_0]^{-1}V|d\rangle$ and
Eqs.~(\ref{Seq1}) and (\ref{Seq2}), we can obtain
\begin{align}
\langle\alpha|d\rangle&=\langle\alpha
|[H-H_0]^{-1}V|d\rangle\nonumber\\
&=-\frac{\langle\alpha|V|d\rangle}{E(\alpha)+B} \ .
\end{align}
Then, Eq.~(\ref{comple2}) can be written as
\begin{equation}\label{ZZ}
1-Z=\int d\alpha\frac{|\langle
\alpha|V|d\rangle|^2}{(E(\alpha)+B)^2}.
\end{equation}
For small $B$, the above integral nearly diverges, so it can then be
approximately evaluated by restricting $|\alpha\rangle$ to
low-energy two-particle states. If the coupling between $|d\rangle$
and the two-particle state is an $S$-wave coupling, we can then
replace $|\langle\alpha|V|d\rangle|$ by $g$, and replace the
$\alpha$ integral with
\begin{equation}
d\alpha=\frac{4\pi
p^2dp}{(2\pi)^3}=\frac{\mu^{3/2}}{\sqrt{2}\pi^2}E^{1/2}dE,\ \ \ \ \
\  E\equiv p^2/2\mu,
\end{equation}
where $\mu$ is the reduced mass of the two constituents. After these
replacements we then obtain the effective coupling constant
\begin{equation}\label{coupling}
g^2=\frac{2\pi\sqrt{2\mu B}}{\mu^2}(1-Z) \ ,
\end{equation}
which encodes the structure information of the composite
system~\cite{Baru:2003qq}. Notice that our defination for $g^2$ has
a relative factor $(2\pi)^3$ compared with that in
Ref.~\cite{Weinberg2}. We use this convention, because this make it
convenient to incorporate the compositeness theorem in the EFT
approach.

\begin{center}
  \vspace*{6pt}\centering
  \includegraphics{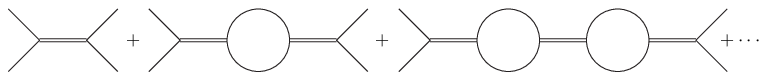}
  \\ [6pt]
  \figcaption{\label{fig} Feynman diagrams for the two particle scattering. The double line denotes the bare state.}
  \end{center}

Now we will incorporate the compositeness theorem in the EFT
approach. Consider a bare state $|\mathcal{B}\rangle$ with bare mass
$-B_0$ and coupling $g_0$ to the two-particle state. If
$|\mathcal{B}\rangle$ is near the two-particle threshold, then the
leading two-particle scattering amplitude can be obtained by summing
the Feynman diagrams in Fig.~\ref{fig}. Later, we will provide a
power counting argument to justify the summation. Near threshold,
the momenta of these two particles are non-relativistic. Therefore,
the loop integral in Fig.~\ref{fig} can be done the same way as that
in Ref.~\cite{KSW1,KSW2}. With the minimal subtraction (MS) scheme,
the result of the loop integral can be written as
\begin{align}
\mathcal{I}^{MS}&\equiv\int\frac{d^D\ell}{(2\pi)^D}\frac{i}{[\ell^0-\vec{\ell}^2/(2m_1)+i\epsilon]}\nonumber\\
&\quad{}\times\frac{i}{[E-\ell^0-\vec{\ell}^2/(2m_2)+i\epsilon]}\nonumber\\
&=i\frac{\mu}{2\pi}\sqrt{-2\mu E-i\epsilon}.
\end{align}
Thus, the Feynman amplitude for Fig.~\ref{fig} reads
\begin{equation}
\mathcal{A}=-\frac{g_0^2}{E+B_0-g_0^2\frac{\mu}{2\pi}\sqrt{-2\mu
E-i\epsilon}} . \label{firstAmp}
\end{equation}
Because a physical bound state $|d\rangle$ corresponds to a pole at
$E=-B$, we have
\begin{equation}
B_0-g_0^2\frac{\mu}{2\pi}\sqrt{2\mu B}\equiv B,\ \Rightarrow
\ B_0=B+g_0^2\frac{\mu}{2\pi}\sqrt{2\mu B}.\label{baremass}
\end{equation}
Then, the amplitude can be written as
\begin{equation}
\mathcal{A}=\frac{\delta^\prime}{E+B+\tilde{\Sigma}^\prime(E)},\label{fullamp}
\end{equation}
where $\delta^\prime=-\frac{g_0^2}{1+g_0^2\mu^2/(2\pi\sqrt{2\mu
B})}$ is the residual of the bound state pole, and
\begin{equation}
\tilde{\Sigma}^\prime(E)=\delta^\prime[\frac{\mu}{2\pi}\sqrt{-2\mu
E-i\epsilon}+\frac{\mu\sqrt{2\mu B}}{4\pi B}(E-B)].
\end{equation}
Since $\delta^\prime$ is the residual of the bound state pole, it
naturally leads to the connection of $\delta^\prime\equiv -g^2$
where $g^2$ is defined in Eq.~(\ref{coupling}) as the effective
coupling constant. Therefore, the leading order amplitude for the
two-particle scattering can be written as
\begin{equation}
\mathcal{A}=-\frac{g^2}{E+B+\tilde{\Sigma}^\prime(E)},\label{finalAmp}
\end{equation}
where $\tilde{\Sigma}^\prime(E)$ can now been written as
\begin{equation}
\tilde{\Sigma}^\prime(E)=-g^2[\frac{\mu}{2\pi}\sqrt{-2\mu
E-i\epsilon}+\frac{\mu\sqrt{2\mu B}}{4\pi B}(E-B)].
\end{equation}

One can easily check that the amplitude given in
Eq.~(\ref{finalAmp}) satisfies the unitary condition. Actually, the
same solution as Eq.~(\ref{finalAmp}) was obtained by Weinberg fifty
years ago, but with a different approach~\cite{Weinberg2}. With
$\delta^\prime=-g^2$ and Eq.~(\ref{coupling}), we obtain
\begin{equation}\label{barecoupling}
g_0^2=\frac{2\pi\sqrt{2\mu B}}{\mu^2}\frac{1-Z}{Z}=g^2/Z.
\end{equation}
Combining Eq.~(\ref{baremass}) and (\ref{barecoupling}) together we
obtain
\begin{equation}\label{finalbaremass}
B_0=\frac{2-Z}{Z}B.
\end{equation}
The limit $B_0\rightarrow \infty$ corresponds to $Z\rightarrow 0$,
which is consistent with the condition discussed
in~\cite{Weinberg1}. Also, Eq.~(\ref{barecoupling}) defines the wave
function renormalization constant of $|\mathcal{B}\rangle$, i.e.
$Z={1}/{[1+g_0^2\mu^2/(2\pi\sqrt{2\mu B})]}$, which is the same as
the result in Ref.~\cite{Martin}. Comparing Eq.~(\ref{finalAmp}) with
the effective range expansion formula
\begin{equation}
\mathcal{A}=\frac{2\pi}{\mu}\frac{1}{-1/a+\frac{1}{2}r_0p^2-ip+\mathcal{O}(p^4)},\label{ERE}
\end{equation}
we have the famous relations which were first obtained by
Weinberg~\cite{Weinberg2}
\begin{equation}
a=\frac{2(1-Z)}{2-Z}\frac{1}{\sqrt{2\mu B}},\ \ \ \ \ \
r_0=-\frac{Z}{1-Z}\frac{1}{\sqrt{2\mu B}},\label{ar}
\end{equation}
where $a$ is the scattering length and $r_0$ is the effective range.

It is interesting to examine the behavior of the tree diagram
amplitude in Fig.~\ref{fig} near threshold. The tree diagram
amplitude reads
\begin{equation}
\mathcal{A}_{tree}=-\frac{g_0^2}{E+B_0}.
\end{equation}
In the limit $Z\rightarrow 0$, we have $B_0\rightarrow \infty$ and
$g_0\rightarrow \infty$. However, $\mathcal{A}_{tree}$ is still well
defined. With Eqs.~(\ref{barecoupling}) and (\ref{finalbaremass}),
we have
\begin{equation}
\lim_{Z\rightarrow 0}\mathcal{A}_{tree}=-\frac{2\pi}{\mu\sqrt{2\mu
B}}. \label{tree}
\end{equation}
Equation~(\ref{tree}) is just the equivalence of the four-Fermi
theory and Yukawa theory found in Ref.~\cite{Lurie}. To be
explicit, in the limit $Z\rightarrow 0$, the $S$-channel resonance
exchange interaction will reduce to the contact interaction. Since
we are interested in the low energy physics, the binding momentum
$\gamma=(2\mu B)^{1/2}$ and the three-momentum of the two-particle state
$p$ is small. We can count these two momenta as the same order, i.e.
$\gamma, \ p\sim\mathcal{O}(p)$. One can see that due to the
existence of the bound state the coefficient of the leading contact
term can be enhanced to the order of $(2\mu
B)^{-1/2}\sim\mathcal{O}(p^{-1})$. Hence in such a case all the
bubble diagrams of the leading contact term are equally important and
should be resummed at the leading order. It is straightforward to
apply this power counting argument to the case $Z\neq0$. One then
find that all the Feynman diagrams in Fig.~\ref{fig} are at the same
order of $\mathcal{O}(p^{-1})$, therefore they should be resummed. A
similar power counting argument to support the summation of the
leading contact term is provided in Ref.~\cite{KSW1,KSW2}, in which
a new subtraction scheme i.e. power divergence subtraction (PDS) scheme,
is proposed. If we use the PDS scheme, then Eq.~(\ref{firstAmp})
will become
\begin{equation}
\mathcal{A}_{PDS}=-\frac{g_0^2}{E+B_0-g_0^2\frac{\mu}{2\pi}(\sqrt{-2\mu
E-i\epsilon}-\Lambda_{PDS})},
\end{equation}
where $\Lambda_{PDS}$ is the dimensional regularization parameter.
By extracting the bare mass similar to what we have done above, we
find that Eqs.~(\ref{finalAmp})--(\ref{barecoupling}) still
hold but Eq.~(\ref{finalbaremass}) will be changed to
\begin{equation}
B_0=\frac{2-Z}{Z}B-\frac{1-Z}{Z}\sqrt{2B/\mu}\Lambda_{PDS}.\label{PDSmass}
\end{equation}
If $Z\neq 0$ and $Z\neq1$, the bare mass $B_0$ determined from
Eq.~(\ref{PDSmass}) will depend on the regularization parameter
$\Lambda_{PDS}$, which can be arbitrary. It means that the
determination of the bare mass will inevitably depend on the scheme
as emphasized in Ref.~\cite{Ronchen:2012eg}. As a consequence, one
presumably need not worry too much about the physical meaning of a
bare mass. In contrast, the physical observable such as
Eq.~(\ref{finalAmp}) is scheme-independent and can be determined by
measuring the line shape.

In the above, we have incorporated the compositeness theorem in the
EFT, and we obtain the leading order amplitude for the low energy
$S$-wave two-particle scattering when a bound state exists, which is
given in Eq.~(\ref{finalAmp}). It is interesting and important to
compare the amplitude with other low energy amplitudes which are
widely used in studying the structure of the $XYZ$ states. In
Refs.~\cite{Braaten:2007dw,Stapleton:2009ey}, the authors use the
following low energy amplitude
\begin{equation}
f(E)=\frac{1}{-1/a+\sqrt{-2\mu E-i\epsilon}},\label{fE}
\end{equation}
One can find that $f(E)$ is just the amplitude given in
Eq.~(\ref{ERE}) if the term $\frac{1}{2}r_0 p^2$ is neglected. From
Eq.~(\ref{ar}), one can find that if $Z=0$, then $r_0=0$, and the
term $\frac{1}{2}r_0 p^2$ disappears in the amplitude. However, if
$Z\neq0$, $\frac{1}{2}r_0 p^2$ is at the order of $\mathcal{O}(p)$,
which is the same as the term $\sqrt{-2\mu E-i\epsilon}$ or $-ip$.
Therefore the term $\frac{1}{2}r_0 p^2$ cannot be neglected in the
low energy amplitude for $Z\neq0$. This suggests that $f(E)$ can
only be used if a bound state is a pure molecule.

In Refs.~\cite{Zhang:2009bv,Hanhart:2007yq,Kalashnikova:2009gt}, the
authors use the Flatt\'{e} parametrization for the low energy
amplitude. Considering only the single channel coupling, the
Flatt\'{e} amplitude reads
\begin{equation}
F(E)=-\frac{1}{2}\frac{g_1}{E-E_f-\frac{1}{2}g_1\sqrt{-2\mu
E-i\epsilon}+i\frac{1}{2}\Gamma}.
\end{equation}
Neglecting $\Gamma$, one can find that $F(E)$ is essentially the
same as our amplitude given in Eq.~(\ref{firstAmp}). By comparing
$F(E)$ with Eq.~(\ref{firstAmp}), one can find that
\begin{equation}
g_1=g_0^2\mu/\pi,\ \ \ \ \ \ \ E_f=-B_0.
\end{equation}
With Eq.~(\ref{barecoupling}) and Eq.~(\ref{finalbaremass}), one can
realize that $F(E)$ can only be applied if $Z$ is not very small, or
the bound state contains a substantial compact component, because in
the limit $Z\rightarrow 0$, the Flatt\'{e} parameters $g_1$ and
$E_f$ become infinite. In such a case the fitting with Flatt\'{e}
parametrization will exhibit scaling behavior as was found in
Ref.~\cite{Hanhart:2007yq}. Therefore, if $Z\rightarrow 0$, in order
to obtain the parameters of the near threshold state, one had better
use $F(E)$ in the form that both the numerator and denominator are
multiplied by a factor $Z$. One can easily find that in such form $F(E)$
is just $f(E)$, which is given in Eq.~(\ref{fE}). In short, $f(E)$
can only be applied if the bound state is purely dynamically
generated or $Z=0$, and $F(E)$ can only be applied if the bound state
contains a substantial compact component. In contrast, the low energy
amplitude given in Eq.~(\ref{finalAmp}) can be used in both
cases, and in the study of the $XYZ$ states, it is better to use the
amplitude given in Eq.~(\ref{finalAmp}).

\section{Study of the $X(3872)$ in the EFT approach}

We show how to incorporate Weinberg's compositeness theorem in the
EFT and apply it to the study of threshold states in which
both elementary and molecular configurations could be present.
Although most of the formulae we present above had been obtained
with the quantum mechanical approach~\cite{Weinberg2,Baru:2003qq},
it is still useful to reproduce them in the EFT approach. The idea
is that with the EFT, we can obtain the relevant Feynman rules for
the near-threshold states. These Feynman rules can then be directly
applied to processes involving such states as a more realistic
prescription for their threshold behaviors. What is more important
is that with the EFT approach, we can set up the power counting and
study the higher order corrections systematically. Therefore the EFT
approach  can be applied much more easily to phenomenological studies and
may provide a clearer physical picture for some of those threshold
states. We also mention that some of those points have been
addressed or demonstrated in the recent analyses of
Refs.~\cite{Cleven:2013sq,Wang:2013cya,Guo:2013zbw}.

In the following, as an application we will use the EFT approach  to
study the structure of the $X(3872)$. Before proceeding, we would
like to clarify that our approach for the $X(3872)$ is different
from the XEFT approach~\cite{XEFT} where the $X(3872)$ is assumed to
be a weakly bound molecule of the $D^{\ast 0}\bar D^0$ pair.  This
corresponds to the special case with $Z=0$ in the EFT approach.
Instead, we do not make any assumption on the structure of the
$X(3872)$ in advance, i.e. we leave the $Z$ as a free parameter
which can be determined by the physical observables.

First, we give the relevant Feynman rules for the $X(3872)$ in our
EFT approach. The propagator of the $X(3872)$ is
\begin{equation}
G(E)_X=\frac{iZ}{E+B+\tilde{\Sigma}^\prime(E)+i\Gamma/2},
\end{equation}
where $\Gamma$ denotes the width of the $X(3872)$ which comes from
the decay modes that do not proceed through its $D^{\ast 0}\bar D^0$
component. Our convention is that a factor of $\sqrt{2M_X}$ has been
absorbed into the field operator of the $X(3872)$. It is convenient
to use this convention for the boson in the nonrelativistic
formalism. Hence a boson field has the dimension of $3/2$ and the
Feynman rule for an external boson should be $\sqrt{2M}$. Actually,
the coupling constant $g^2$ in Eq.~(\ref{coupling}) is defined under
this convention. The Feynman rule for the $XD^{\ast 0}\bar D^0$
coupling is given as
\begin{equation}
i\frac{g_0}{\sqrt{2}}=i\left(\frac{g^2}{2Z}\right)^{1/2},
\end{equation}
where the factor $1/\sqrt{2}$ is due to the definition of the
$C$-even state $(D^{\ast 0}\bar D^0+D^0\bar
D^{\ast 0})/\sqrt{2}$.

As mentioned before, near threshold, there are two small momenta,
the binding momentum $\gamma=(2\mu B)^{1/2}$ and the three momentum
of the charmed meson $p$. We can count these two momenta as the same
order, i.e. $\gamma, \ p\sim\mathcal{O}(p)$. Therefore, we can find
that $E, \ B\sim\mathcal{O}(p^2)$ and $g\sim \mathcal{O}(p^{1/2})$.
One can then easily check that the elastic scattering amplitude
given in Eq.~(\ref{finalAmp}) is at the order of
$\mathcal{O}(p^{-1})$ which is consistent with the result in
Ref.~\cite{KSW1,KSW2}.

Now we come to describe the line shape of $D^{\ast 0}\bar D^0$ in
$B\rightarrow X(3872) K \rightarrow D^{\ast 0}\bar D^0 K $ in this
EFT approach. For studies of the line shape with other approaches
one can refer to
Refs.~\cite{Zhang:2009bv,Hanhart:2007yq,Kalashnikova:2009gt,Braaten:2007dw,Stapleton:2009ey}.
In Ref.~\cite{Braaten:2007dw,Stapleton:2009ey}, the authors use the
amplitude $f(E)$ which is given in Eq.~(\ref{fE}) in their analysis.
As we have mentioned before, $f(E)$ can only be used for a pure
molecule. However, a pure molecule assignment for the $X(3872)$
seems to conflict with the recent LHCb
measurement~\cite{Aaij:2014ala}. In
Refs.~\cite{Zhang:2009bv,Hanhart:2007yq,Kalashnikova:2009gt}, the
authors describe the $D^{\ast 0}\bar D^0$ line shape with Flatt\'e
parametrization. Assuming the $X(3872)$ production via the
short-distance process, Ref.~\cite{Kalashnikova:2009gt} further
addresses the question of a possible $\chi^\prime_{c1}$ charmonium
admixture in the wave function of the $X(3872)$. The idea is to
integrate the spectral density which can be expressed in terms of
Flatt\'{e} parameters. However, in this approach the integration
bounds are somewhat arbitrary, hence this approach is inevitably
model dependent. Comparing with these approaches, the advantage of
the our approach is that we make no assumptions on the structure of
the $X(3872)$ in advance. In this way, one can then clearly address
the question whether a bound state is a pure molecule or it contains a
substantial compact component. Another advantage of our approach is
that instead of making assumptions on the production mechanism of
the $X(3872)$ as in
Ref.~\cite{Zhang:2009bv,Hanhart:2007yq,Kalashnikova:2009gt}, we
systematically consider both the short and long-distance production
mechanisms of the $X(3872)$. In the short-distance production
mechanism the $X(3872)$ is produced directly at the short-distance
vertex of the $B$ decay, while in the long-distance production
mechanism a $D^{\ast 0}\bar D^0$ pair is produced first in the $B$
decay and then rescatters  into the $X(3872)$. The answer to the
question about which production mechanism is more important than the
other would depend on the structure of the $X(3872)$. As follows,
instead of making assumptions on the structure of the $X(3872)$ in
advance, we actually consider both these different production
mechanisms in our analysis.

The leading order Feynman diagrams for these two different
production mechanisms are presented in Fig.~\ref{fig2}, for which
the Feynman amplitudes can be explicitly expressed as
\begin{align}
i\mathcal{M}_a&=-\frac{\mathcal{A}_{XK}}{\sqrt{2}}\frac{\sqrt{Z}g}{E+B+\tilde{\Sigma}^\prime(E)+i\Gamma/2}\vec{p}_K\cdot\vec{\epsilon}^\ast,                         \nonumber\\
i\mathcal{M}_b&=\mathcal{B}_{DDK}\frac{\mu}{2\pi}\frac{g^2\sqrt{-2\mu
E-i\epsilon}}{E+B+\tilde{\Sigma}^\prime(E)+i\Gamma/2}\vec{p}_K\cdot\vec{\epsilon}^\ast,\label{Amp}
\end{align}
where $p_K$ is the momentum of the $K$ meson in the rest frame of
the $B$ meson, and $\epsilon$ is the polarization vector of the
outgoing $D^{\ast 0}$. We use $\mathcal{A}_{XK}$ and
$\mathcal{B}_{DDK}$ to denote the first production vertices in
Fig.~\ref{fig2}, i.e. $B\rightarrow X(3872)K$ and $B\rightarrow
D^{\ast 0}\bar D^0 K$, respectively. Near the threshold of $D^{\ast
0}\bar D^0$, we can treat $\mathcal{A}_{XK}$ and $\mathcal{B}_{DDK}$
as constants. Note that we have omitted the factors  from the
external $D^{(\ast)}$ mesons in Eq.~(\ref{Amp}) which can be
absorbed into $\mathcal{A}_{XK}$ and $\mathcal{B}_{DDK}$.

\begin{center}
  \includegraphics{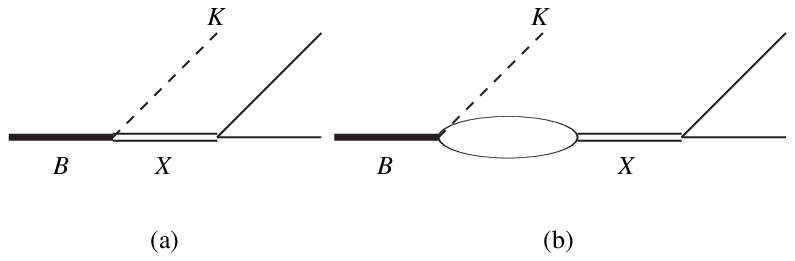}\\
  \figcaption{\label{fig2} Feynman diagrams for $B\rightarrow X(3872)K\rightarrow D^{\ast 0}  \bar D^0 K$. Solid lines in the
  loop and final state represent the charm and anti-charm mesons.}
  \end{center}

It is easy to count the power of the above amplitudes and one can
find $\mathcal{M}_a\sim\mathcal{O}(p^{-3/2})$ and
$\mathcal{M}_b\sim\mathcal{O}(p^0)$. From the power counting, one
may find that the short-distance production mechanism is more
important than the long-distance one. However, it should be noted
that $\mathcal{M}_a$ is proportional to the factor $\sqrt{Z}$.
Therefore, its contribution will be suppressed if the $X(3872)$ is
dominated by a molecular component. It is interesting to note that
with $Z=0$ the term of $\mathcal{M}_a$ will vanish, and then the
production of the $X(3872)$ will only come from the long-distance
production mechanism $\mathcal{M}_b$. This feature ensures that our
separation of the short-distance production mechanism from the
long-distance one makes sense.

Taking into account the  non-resonance production contribution, the
full amplitude to describe $B\rightarrow D^{\ast 0}\bar D^0 K$ is
expressed as
\begin{equation}
i\mathcal{M}=i\mathcal{M}_a+i\mathcal{M}_b+\mathcal{B}_{DDK}\vec{p}_K\cdot\vec{\epsilon}^\ast(D^\ast)
\ ,
\end{equation}
where the term
$\mathcal{B}_{DDK}\vec{p}_K\cdot\vec{\epsilon}^\ast(D^\ast)$
describes the non-resonance production which is at the same order as
$i\mathcal{M}_b$. Now we can use the above amplitude to describe the
Belle and BaBar data~\cite{Bfactory1,Bfactory2}. The free parameters
in our calculation include $\Gamma$, $B$, $Z$, $\mathcal{A}_{XK}$
and $\mathcal{B}_{DDK}$. However, the experimental data have
large error bars. To reduce the uncertainty, we fix $\Gamma$
and $B$ with the values that are determined in $X(3872)\rightarrow
J/\psi X$, where $X$ denotes the light hadrons. The reason is
because the data from the decay modes of $X(3872)\rightarrow J/\psi
X$ have higher statistics and there the $X(3872)$ appears as a
narrow Breit-Wigner structure. We adapt the PDG~\cite{PDG} value
$M_{X(3872)}=3871.68$ MeV for the mass of $X(3872)$, which is the
average over the measurements from the decay modes of
$X(3872)\rightarrow J/\psi X$. With $M_{D^{\ast 0}}=2006.99$ MeV and
$M_{\bar D^0}=1864.86$ MeV~\cite{PDG}, we can fix the binding energy
as $B=0.17$ MeV. The width of $X(3872)$ is not settled by the
data for $X(3872)\rightarrow J/\psi X$, but the upper limit is given
as $\Gamma<1.2$ MeV. Since the width is small, we fix the
non-$D^{\ast 0}\bar D^0$ width $\Gamma=0$ in our fitting. We have
checked that the $D^{\ast 0}\bar D^0$ line shape is not sensitive to
$B$ and $\Gamma$ around the fixed values. Therefore, our fitting
parameters are $\mathcal{A}_{XK}$, $\mathcal{B}_{DDK}$ and $Z$.

\begin{center}
\includegraphics{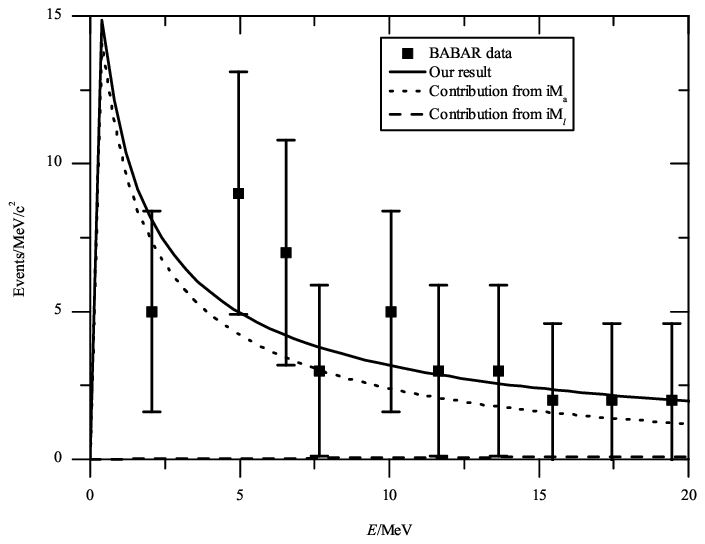}\ \ \ \ \ \ \ \ \includegraphics{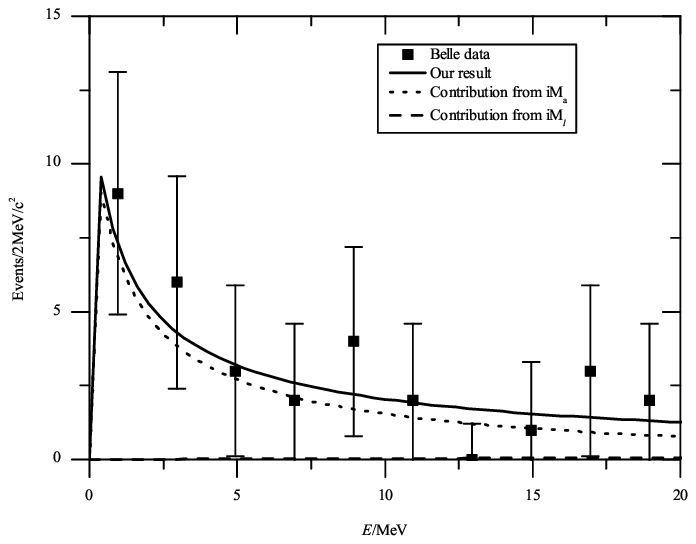}
\figcaption{\label{DATA}  The line shape of the $D^{\ast 0}\bar D^0$
spectrum in $B\rightarrow D^{\ast 0}\bar D^0 K$. The data are from
Refs.~\cite{Bfactory1,Bfactory2}. The solid line denotes the overall
fitting result, the dotted line is the contribution from
$i\mathcal{M}_a$, and the dashed line that from
$i\mathcal{M}_\ell=i\mathcal{M}_b+\mathcal{B}_{DDK}\vec{p}_K\cdot
\vec{\epsilon}^\ast(D^\ast)$. An arbitrary normalization is
implemented.}
  \end{center}

The fitting results are presented in Fig.~\ref{DATA} and compared
with the experimental data~\cite{Bfactory1,Bfactory2}. Notice that
there is an arbitrary scaling factor between the BABAR and Belle
data; we fit the ratio ${\mathcal{A}_{XK}}/{\mathcal{B}_{DDK}}$ and
$Z$ for these two sets of data simultaneously but leave a free scale
factor to be fitted by the data. This, in principle,
introduces an additional parameter and leads to  $\chi^2/d.o.f=0.4$
which indicates some correlations among the parameters. This can be
improved by future experimental measurement. For the physical
discussion, we only list the fitted ratio
${\mathcal{A}_{XK}}/{\mathcal{B}_{DDK}}$ and parameter $Z$ as
follows
\begin{equation}
\frac{\mathcal{A}_{XK}}{\mathcal{B}_{DDK}}=-0.15\pm 0.65 \
\mbox{GeV}^{3/2},\ \ \ \ \ Z=0.19\pm 0.29.\label{result}
\end{equation}

In Fig.~\ref{DATA}, we also show the contribution from different
pieces of the amplitude, i.e. $i\mathcal{M}_a$ and
$i\mathcal{M}_\ell=i\mathcal{M}_b+\mathcal{B}_{DDK}\vec{p}_{K}\cdot\vec{\epsilon}^\ast(D^\ast)$
as the dotted and dashed line, respectively. From Eq.~(\ref{result})
one can see that the fitted parameters are with large uncertainties
due to the large error bars with the BABAR and Belle data. To reduce
the number of the free parameters one can only use the leading order
amplitude $i\mathcal{M}_a$ in the fitting. The fitted $Z$ is
$Z=0.12\pm0.11$. The fitting quality is almost the same as that of
Eq.~(\ref{result}). Because the fitting line shape is similar to that
in Fig.~\ref{DATA}, we will not bother to show it again.

Due to the relatively large uncertainties with the fitted
parameters, we discuss the following possible scenarios arising from
the fitting results:

\begin{itemize}

\item The main feature of Fig.~\ref{DATA} is that a small nonvanishing value of $Z$ will result in a sizeable contribution from
the short-distance process, i.e. $i\mathcal{M}_a$. This indicates
that the production of the $X(3872)$ in the $B$ decay is driven by
the short-distance production mechanism. Even a small component of
the $c\bar{c}$ core will lead to a relatively larger production rate
for the $X(3872)$ in comparison with when it is treated as a pure $D^{\ast
0}\bar D^0$ molecule.  Nevertheless, the dominance of the
short-distance production mechanism seems to always produce the
threshold enhancement which may bring concerns about the molecular
feature of the $X(3872)$. However, this may provide a natural
explanation for the sizeable production rate for the $X(3872)$ in
the $B$ decay, and also explain the large isospin violations given
that the compact $c\bar{c}$ component can couple strongly to the
charged $D^*\bar D+c.c.$ pair. This will give rise to enhanced
isospin violation transitions into $J/\psi\rho$ via the intermediate
charged and neutral $D$ meson loops as discussed in the literature.
If the compact component of the $X(3872)$ is $\chi_{c1}^\prime$, its
production rate in the $B$ decay should be comparable with that of
$\chi_{c1}$~\cite{charm1}. Meanwhile, if the $X(3872)$ is a pure
molecule, its production rate will be strongly suppressed. The
recent PDG result gives $\mbox{Br}(B^+\rightarrow \chi_{c1}
K^+)=(4.79\pm0.23)\times 10^{-4}$, while the production ratio of the
$X(3872)$ is constrained as $\mbox{Br}(B^+\rightarrow X(3872)
K^+)<3.2\times 10^{-4}$~\cite{PDG}. Thus, it is not conclusive for
the structure of the $X(3872)$ based on such a measurement. We
expect that a more precise measurement of the decay rate of $B\to
X(3872) K$ would provide a quantitative constraint on the $X(3872)$
structure in the future.

\begin{center}
\includegraphics{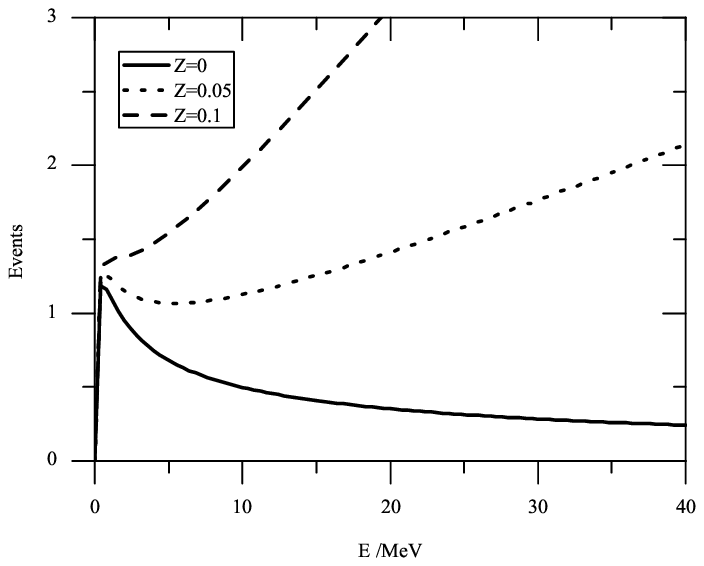}
\figcaption{\label{ZZ-figure} The exclusive contribution from
$i\mathcal{M}_\ell$ to the line shape of the $D^{\ast 0}\bar D^0$
spectrum in $B\rightarrow D^{\ast 0}\bar D^0 K$ with different $Z$.
Here, we set $B=0.5$~MeV as an illustration. The results with
$B=0.17$~MeV are similar. An arbitrary normalization is
implemented.}
\end{center}

\item
It is interesting to discuss the behavior of the term
$i\mathcal{M}_\ell$ in the line shape of $D^{\ast 0}\bar D^0$. In the
case where the $X(3872)$ is a pure molecule, i.e. $Z=0$, the line
shape will be determined by $i\mathcal{M}_\ell$ with
$i\mathcal{M}_a=0$. For convenience, we can express
$i\mathcal{M}_\ell$ by a more compact form
\begin{align}
i\mathcal{M}_\ell&=i\mathcal{M}_b+\mathcal{B}_{DDK}\vec{p}_K\cdot\vec{\epsilon}^\ast(D^\ast)\nonumber\\
&=\mathcal{B}_{DDK}\frac{ZE+(2-Z)B}{E+B+\tilde{\Sigma}^\prime(E)}\vec{p}_K\cdot\vec{\epsilon}^\ast(D^\ast),\label{Zdependent}
\end{align}
where we have set $\Gamma=0$ as discussed above. By setting $Z=0$
the energy dependence of $\mathcal{M}_\ell$ is just the same as
$f(E)$, which is used in Ref
.~\cite{Braaten:2007dw,Stapleton:2009ey}. As discussed before, if we
fix $Z=0$, these two approaches should indeed converge as expected.
However, the explicit $Z$-dependence will bring novel aspects to the
line shape description.

We can take a closer look at the $Z$-dependence of
$i\mathcal{M}_\ell$, which is illustrated by  the line shape in
Fig.~\ref{ZZ-figure}. Note that  Fig.~\ref{ZZ-figure} is rescaled by
an arbitrary factor due to the unknown value of the cross sections.
One can see that the line shape is very sensitive to $Z$. If $Z=0$,
the line shape has a clear near-threshold enhancement. But when $Z$
increases, the near-threshold enhancement disappears quickly.  The
reason is that if $Z\neq 0$ the factor $ZE$ in the numerator of
Eq.~(\ref{Zdependent}) will play an important role in the line
shape. From Eq.~(\ref{Zdependent}) one can also find that, for
$Z=0$, $i\mathcal{M}_\ell$ is proportional to the small binding
energy $B$. Therefore, we can conclude that if $Z$ is
non-negligible, the near-threshold enhancement of the $D^{\ast
0}\bar D^{0}$ mass spectrum in the $B$ decay will be driven by the
short-distance production mechanism of the $X(3872)$, although the
dominant component of the $X(3872)$ is molecular. This feature is
again consistent with the success of treating the $X(3872)$ as pure
molecule in the explanation of the line
shape~\cite{Braaten:2007dw,Stapleton:2009ey}. One should note that
even if the long-distance production is enhanced by some unexpected
mechanism, the conclusion is still true.

\item
One may consider to further describe the line shape measured from
$X(3872)\rightarrow J/\psi
\pi^+\pi^-$~\cite{Aubert:2008gu,Adachi:2008te} in order to have a
better determination of $Z$. However, since the coupling between
$X(3872)$ and $J/\psi\pi^+\pi^-$ is unclear, namely, they may couple
directly or through the intermediate meson loops, the inclusion of
the line shape measured in $X(3872)\rightarrow J/\psi \pi^+\pi^-$
will inevitably introduce more free parameters. This will be studied
in the future with the availability of more precise experimental
data.

\item
In obtaining the above amplitudes, the MS scheme is adopted to
evaluate the loop integral. It is still interesting to discuss the
results when the PDS scheme is adopted. With the PDS scheme, the
amplitude $\mathcal{M}_a$ remains the same but the amplitude
$\mathcal{M}_\ell$ will change to
\begin{align}
i\mathcal{M}_\ell
&=\mathcal{B}_{DDK}\times \vec{p}_K\cdot\vec{\epsilon}^\ast\times\nonumber\\
&{}\frac{ZE+(2-Z)B-(1-Z)\sqrt{2B/\mu}\Lambda_{PDS}}{E+B+\tilde{\Sigma}^\prime(E)}
. \label{PDSMl}
\end{align}
If $Z=0$, the arbitrary scale $\Lambda_{PDS}$ can be absorbed into
the definition of $\mathcal{B}_{DDK}$ to make sure that the physical
amplitude $\mathcal{M}_\ell$ does not depend on this arbitrary
scale. However, if $0<Z<1$, it seems impossible to do that due to
the factor $ZE$ in the numerator. Therefore, for $0<Z<1$ the
amplitude $\mathcal{M}_\ell$ will inevitably depend on the arbitrary
scale $\Lambda_{PDS}$ if the PDS scheme is adopted. Whether this
means that the PDS scheme may not be suitable for the study of the
decay processes in our EFT needs to be further investigated. We note
that the same problem does not occur in two-particle elastic
scattering in the EFT approach as discussed before.

\end{itemize}

\section{Summary}

In summary, we have proposed an EFT approach with the compositeness
theorem incorporated for the study of threshold states. By
determining $Z$, which is the probability of finding an elementary
component in the bound state via physical observables, our EFT
approach can be used to identify the structure of the $S$-wave
near-threshold states. As an example of the application, we use the
EFT approach to describe the line shape of the $D^{\ast 0}\bar D^0$
mass spectrum in the decay of $B\rightarrow D^{\ast 0}\bar D^0 K$.
By fitting the data from BaBar and Belle, we obtain a nonvanishing value
of $Z=0.19\pm 0.29$. Although higher statistics data for
$B\rightarrow X(3872)(\rightarrow D^{\ast 0}\bar D^0)K$ are needed
to reduce the uncertainty of $Z$, the study of the $Z$ dependence of
the transition amplitudes suggests that a small value of $Z$ inside
the $X(3872)$ would cause the threshold enhancement in the $D^{\ast
0}\bar D^0$ invariant mass spectrum via the short-distance
production mechanism. It alternatively implies that the $X(3872)$ is
dominated by the molecular $D^{\ast 0}\bar D^0$ molecular component.
This scenario can naturally explain the observation of sizeable
isospin violation decays of $X(3872)\to J/\psi\rho^0$ via the
charged and neutral $D^{\ast}\bar D+c.c.$ meson loops as the leading
contribution. Finally, it will be very interesting to constrain $Z$
from other approaches. For example, in Ref.~\cite{Meng:2013gga} the
determined value of $Z$ is $Z=(28-44)\%$, which is close to our
result.

\acknowledgments{We would like to thank Hanqing Zheng and Christoph
Hanhart for valuable discussions and suggestions.}

\end{multicols}

\vspace{10mm}

\vspace{-1mm}
\centerline{\rule{80mm}{0.1pt}}
\vspace{2mm}

\begin{multicols}{2}

\end{multicols}

\clearpage

\end{CJK*}

\begin{thebibliography}{90}

\vspace{3mm}

%\cite{Belle:2011aa}
\bibitem{Belle:2011aa}
  Bondar A et al. (Belle Collaboration).
  %``Observation of two charged bottomonium-like resonances in Y(5S) decays,''
  Phys.\ Rev.\ Lett., 2012, \textbf{108}: 122001
  %[arXiv:1110.2251 [hep-ex]].
  %%CITATION = ARXIV:1110.2251;%%
  %81 citations counted in INSPIRE as of 11 Jul 2013


%\cite{Ablikim:2013mio}
\bibitem{Ablikim:2013mio}
  Ablikim M et al. (BESIII Collaboration).
  %``Observation of a Charged Charmoniumlike Structure in $e^+e^-$ ¡ú $¦Ð^+¦Ð^-$ J/¦× at $\sqrt{s}$ =4.26????GeV,''
  Phys.\ Rev.\ Lett., 2013, \textbf{110}: 252001
  %[arXiv:1303.5949 [hep-ex]].
  %%CITATION = ARXIV:1303.5949;%%
  %201 citations counted in INSPIRE as of 18 Nov 2014


%\cite{Liu:2013dau}
\bibitem{Liu:2013dau}
  Liu Z Q et al. (Belle Collaboration).
  %``Study of $e^+e^- ¡ú ¦Ð^+ ¦Ð^- J/¦×$ and Observation of a Charged Charmoniumlike State at Belle,''
  Phys.\ Rev.\ Lett., 2013, \textbf{110}: 252002
  %[arXiv:1304.0121 [hep-ex]].
  %%CITATION = ARXIV:1304.0121;%%
  %169 citations counted in INSPIRE as of 18 Nov 2014


%\cite{Xiao:2013iha}
\bibitem{Xiao:2013iha}
  Xiao T, Dobbs S, Tomaradze A et al.
  %``Observation of the Charged Hadron $Z_c^{\pm}(3900)$ and Evidence for the Neutral $Z_c^0(3900)$ in $e^+e^-\to \pi\pi J/\psi$ at $\sqrt{s}=4170$ MeV,''
  Phys.\ Lett.\ B, 2013, \textbf{727}: 366
  %[arXiv:1304.3036 [hep-ex]].
  %%CITATION = ARXIV:1304.3036;%%
  %101 citations counted in INSPIRE as of 18 Nov 2014


%\cite{Ablikim:2013wzq}
\bibitem{Ablikim:2013wzq}
  Ablikim M et al. (BESIII Collaboration).
  %``Observation of a Charged Charmoniumlike Structure $Z_c$(4020) and Search for the $Z_c$(3900) in $e^+e^- \to ¦Ð^+¦Ð^-h_c$,''
  Phys.\ Rev.\ Lett., 2013, \textbf{111}: 242001
  %[arXiv:1309.1896 [hep-ex]].
  %%CITATION = ARXIV:1309.1896;%%
  %78 citations counted in INSPIRE as of 18 Nov 2014


%\cite{Ablikim:2013xfr}
\bibitem{Ablikim:2013xfr}
  Ablikim M et al.\ (BESIII Collaboration).
  %``Observation of a charged $(D\bar{D}^{*})^\pm$ mass peak in $e^{+}e^{-} \to \pi D\bar{D}^{*}$ at $\sqrt{s} =$ 4.26 GeV,''
  Phys.\ Rev.\ Lett., 2014, \textbf{112}: 022001
  %[arXiv:1310.1163 [hep-ex]].
  %%CITATION = ARXIV:1310.1163;%%
  %43 citations counted in INSPIRE as of 18 Nov 2014


\bibitem{Salam}
  Salam A.
  %``Lagrangian theory of composite particles,''
  Nuovo Cim., 1962, \textbf{25}: 224
  %%CITATION = NUCIA,25,224;%%
  %158 citations counted in INSPIRE as of 02 Jan 2014


\bibitem{Weinberg1}
  Weinberg S.
  %``Elementary particle theory of composite particles,''
  Phys.\ Rev., 1963, \textbf{130}: 776
  %%CITATION = PHRVA,130,776;%%
  %296 citations counted in INSPIRE as of 02 Jan 2014


\bibitem{Weinberg2}
  Weinberg S.
  %``Evidence That the Deuteron Is Not an Elementary Particle,''
  Phys.\ Rev., 1965, \textbf{137}: B672
  %%CITATION = PHRVA,137,B672;%%
  %90 citations counted in INSPIRE as of 03 Jan 2014


\bibitem{Lurie}
     Lurie D, Macfarlane A J.  Phys.\ Rev., 1964, \textbf{136}: B816


%\cite{Brambilla:2010cs}
\bibitem{Brambilla:2010cs}
  Brambilla N, Eidelman S, Heltsley B K et al.
  %``Heavy quarkonium: progress, puzzles, and opportunities,''
  Eur.\ Phys.\ J.\ C, 2011, \textbf{71}: 1534
  %[arXiv:1010.5827 [hep-ph]].


\bibitem{Choi:2003ue}
  Choi S K et al. (Belle Collaboration).
  %``Observation of a narrow charmonium - like state in exclusive B+- ---> K+- pi+ pi- J / psi decays,''
  Phys.\ Rev.\ Lett., 2003, \textbf{91}: 262001
  %[hep-ex/0309032].
  %%CITATION = HEP-EX/0309032;%%
  %844 citations counted in INSPIRE as of 03 Jan 2014


\bibitem{Tornqvist:2004qy}
  Tornqvist N A.
  %``Isospin breaking of the narrow charmonium state of Belle at 3872-MeV as a deuson,''
  Phys.\ Lett.\ B, 2004, \textbf{590}: 209
  %[hep-ph/0402237].
  %%CITATION = HEP-PH/0402237;%%
  %306 citations counted in INSPIRE as of 03 Jan 2014


%\cite{Close:2003sg}
\bibitem{Close:2003sg}
  Close F E, Page P R.
  %``The D*0 anti-D0 threshold resonance,''
  Phys.\ Lett.\ B, 2004, \textbf{578}: 119
  %[hep-ph/0309253].
  %%CITATION = HEP-PH/0309253;%%
  %264 citations counted in INSPIRE as of 03 Jan 2014


%\cite{Wong:2003xk}
\bibitem{Wong:2003xk}
  Wong C Y.
  %``Molecular states of heavy quark mesons,''
  Phys.\ Rev.\ C, 2004, \textbf{69}: 055202
  %[hep-ph/0311088].
  %%CITATION = HEP-PH/0311088;%%
  %145 citations counted in INSPIRE as of 03 Jan 2014


%\cite{Braaten:2003he}
\bibitem{Braaten:2003he}
  Braaten E, Kusunoki M.
  %``Low-energy universality and the new charmonium resonance at 3870-MeV,''
  Phys.\ Rev.\ D, 2004, \textbf{69}: 074005
  %[hep-ph/0311147].
  %%CITATION = HEP-PH/0311147;%%
  %149 citations counted in INSPIRE as of 03 Jan 2014


%\cite{Voloshin:2003nt}
\bibitem{Voloshin:2003nt}
  Voloshin M B.
  %``Interference and binding effects in decays of possible molecular component of X(3872),''
  Phys.\ Lett.\ B, 2004, \textbf{579}: 316
  %[hep-ph/0309307].
  %%CITATION = HEP-PH/0309307;%%
  %198 citations counted in INSPIRE as of 03 Jan 2014


%\cite{Swanson:2003tb}
\bibitem{Swanson:2003tb}
  Swanson E S.
  %``Short range structure in the X(3872),''
  Phys.\ Lett.\ B, 2004, \textbf{588}: 189
  %[hep-ph/0311229].
  %%CITATION = HEP-PH/0311229;%%
  %326 citations counted in INSPIRE as of 03 Jan 2014


%\cite{Swanson:2004pp}
\bibitem{Swanson:2004pp}
  Swanson E S.
  %``Diagnostic decays of the X(3872),''
  Phys.\ Lett.\ B, 2004, \textbf{598}: 197
  %[hep-ph/0406080].
  %%CITATION = HEP-PH/0406080;%%
  %138 citations counted in INSPIRE as of 03 Jan 2014


%\cite{Meng:2005er}
\bibitem{charm1}
  Meng C, Gao Y J, Chao K T.
  %``B ---> chi(c1) (1P,2P)K decays in QCD factorization and X(3872),''
  Phys.\ Rev.\ D, 2013, \textbf{87}: 074035
  %[hep-ph/0506222].
  %%CITATION = HEP-PH/0506222;%%
  %41 citations counted in INSPIRE as of 18 Nov 2014


%\cite{Suzuki:2005ha}
\bibitem{charm2}
  Suzuki M.
  %``The X(3872) boson: Molecule or charmonium,''
  Phys.\ Rev.\ D, 2005, \textbf{72}: 114013
  %[hep-ph/0508258].
  %%CITATION = HEP-PH/0508258;%%
  %108 citations counted in INSPIRE as of 03 Jan 2014


%\cite{Bignamini:2009sk}
\bibitem{charm3}
  Bignamini C, Grinstein B, Piccinini F et al.
  %``Is the X(3872) Production Cross Section at Tevatron Compatible with a Hadron Molecule Interpretation?,''
  Phys.\ Rev.\ Lett., 2009,  \textbf{103}: 162001
  %[arXiv:0906.0882 [hep-ph]].
  %%CITATION = ARXIV:0906.0882;%%
  %58 citations counted in INSPIRE as of 03 Jan 2014


%\cite{Danilkin:2010cc}
\bibitem{charm4}
  Danilkin I V, Simonov Y A.
  %``Dynamical origin and the pole structure of X(3872),''
  Phys.\ Rev.\ Lett., 2010, \textbf{105}: 102002
  %[arXiv:1006.0211 [hep-ph]].
  %%CITATION = ARXIV:1006.0211;%%
  %23 citations counted in INSPIRE as of 03 Jan 2014


\bibitem{Prelovsek:2013cra}
  Prelovsek S, Leskovec L.
  %``Evidence for X(3872) from DD* scattering on the lattice,''
  Phys.\ Rev.\ Lett., 2013, \textbf{111}: 192001
  %[arXiv:1307.5172 [hep-lat]].
  %%CITATION = ARXIV:1307.5172;%%
  %8 citations counted in INSPIRE as of 27 Nov 2013


%\cite{Wang:2013kva}
\bibitem{Wang:2013kva}
  Wang P, Wang X G.
  %``Study on X(3872) from effective field theory with pion exchange interaction,''
  Phys.\ Rev.\ Lett., 2013, \textbf{111}: 042002
  %[arXiv:1304.0846 [hep-ph]].
  %%CITATION = ARXIV:1304.0846;%%
  %6 citations counted in INSPIRE as of 18 Nov 2014


%\cite{Baru:2013rta}
\bibitem{Baru:2013rta}
  Baru V, Epelbaum E, Filin A A et al.
  %``Quark mass dependence of the X(3872) binding energy,''
  Phys.\ Lett.\ B, 2013, \textbf{726}: 537
  %[arXiv:1306.4108 [hep-ph]].
  %%CITATION = ARXIV:1306.4108;%%
  %2 citations counted in INSPIRE as of 27 Nov 2013

%\cite{Jansen:2013cba}
\bibitem{Jansen:2013cba}
  Jansen M, Hammer H W, Jia Y.
  %``Light quark mass dependence of the X(3872) in an effective field theory,''
  Phys.\ Rev.\ D, 2014, \textbf{89}: 014033
  %[arXiv:1310.6937 [hep-ph]].
  %%CITATION = ARXIV:1310.6937;%%
  %3 citations counted in INSPIRE as of 18 Nov 2014


%\cite{Aaij:2014ala}
\bibitem{Aaij:2014ala}
  Aaij R et al. (LHCb Collaboration).
  %``Evidence for the decay $X(3872)\rightarrow\psi(2S)\gamma$,''
  Nucl.\ Phys.\ B, 2014, \textbf{886}: 665
  %[arXiv:1404.0275 [hep-ex]].
  %%CITATION = ARXIV:1404.0275;%%
  %17 citations counted in INSPIRE as of 18 Nov 2014


%\cite{Dong:2009uf}
\bibitem{Dong:2009uf}
  Dong Y, Faessler A, Gutsche T et al.
  %``J/psi gamma and psi(2S) gamma decay modes of the X(3872),''
  J.\ Phys.\ G, 2011, \textbf{38}: 015001
  %[arXiv:0909.0380 [hep-ph]].
  %%CITATION = ARXIV:0909.0380;%%
  %26 citations counted in INSPIRE as of 14 Apr 2014


\bibitem{ZhengHan}
  Meng C, Sanz-Cillero J J, Shi M et al.
  %``A Refined Analysis on the $X(3872)$ Resonance,''
  arXiv:hep-ph/1411.3106
  %%CITATION = ARXIV:1411.3106;%%


\bibitem{Baru:2003qq}
  Baru V, Haidenbauer J, Hanhart C et al.
  %``Evidence that the a(0)(980) and f(0)(980) are not elementary particles,''
  Phys.\ Lett.\ B, 2004, \textbf{586}: 53
  %[hep-ph/0308129].
  %%CITATION = HEP-PH/0308129;%%
  %152 citations counted in INSPIRE as of 09 Aug 2013


%\cite{Kaplan:1998tg}
\bibitem{KSW1}
  Kaplan D B, Savage M J, Wise M B.
  %``A New expansion for nucleon-nucleon interactions,''
  Phys.\ Lett.\ B, 1998, \textbf{424}: 390
  %[nucl-th/9801034].
  %%CITATION = NUCL-TH/9801034;%%
  %439 citations counted in INSPIRE as of 03 Jan 2014


%\cite{Kaplan:1998we}
\bibitem{KSW2}
  Kaplan D B, Savage M J, Wise M B.
  %``Two nucleon systems from effective field theory,''
  Nucl.\ Phys.\ B, 1998, \textbf{534}: 329
  %[nucl-th/9802075].
  %%CITATION = NUCL-TH/9802075;%%
  %440 citations counted in INSPIRE as of 03 Jan 2014


%\cite{Cleven:2011gp}
\bibitem{Martin}
  Cleven M, Guo F K, Hanhart C et al.
  %``Bound state nature of the exotic $Z_b$ states,''
  Eur.\ Phys.\ J.\ A, 2011, \textbf{47}: 120
  %[arXiv:1107.0254 [hep-ph]].
  %%CITATION = ARXIV:1107.0254;%%
  %33 citations counted in INSPIRE as of 10 Sep 2013


\bibitem{Ronchen:2012eg}
  Ronchen D, Doring M, Huang F et al.
  %``Coupled-channel dynamics in the reactions piN --> piN, etaN, KLambda, KSigma,''
  Eur.\ Phys.\ J.\ A, 2013, \textbf{49}: 44
  %[arXiv:1211.6998 [nucl-th]].
  %%CITATION = ARXIV:1211.6998;%%
  %8 citations counted in INSPIRE as of 22 Aug 2013


%\cite{Braaten:2007dw}
\bibitem{Braaten:2007dw}
  Braaten E, Lu M.
  %``Line shapes of the X(3872),''
  Phys.\ Rev.\ D, 2007, \textbf{76}: 094028
  %[arXiv:0709.2697 [hep-ph]].
  %%CITATION = ARXIV:0709.2697;%%
  %69 citations counted in INSPIRE as of 03 Jan 2014


%\cite{Stapleton:2009ey}
\bibitem{Stapleton:2009ey}
  Braaten E, Stapleton J.
  %``Analysis of J/psi pi+ pi- and D0 anti-D0 pi0 Decays of the X(3872),''
  Phys.\ Rev.\ D, 2010, \textbf{81}: 014019
  %[arXiv:0907.3167 [hep-ph]].
  %%CITATION = ARXIV:0907.3167;%%
  %29 citations counted in INSPIRE as of 03 Jan 2014


%\cite{Zhang:2009bv}
\bibitem{Zhang:2009bv}
  Zhang O, Meng C, Zheng H Q.
  %``Ambiversion of X(3872),''
  Phys.\ Lett.\ B, 2009, \textbf{680}: 453
  %[arXiv:0901.1553 [hep-ph]].
  %%CITATION = ARXIV:0901.1553;%%
  %21 citations counted in INSPIRE as of 03 Jan 2014

%\cite{Hanhart:2007yq}
\bibitem{Hanhart:2007yq}
  Hanhart C, Kalashnikova Y S, Kudryavtsev A E et al.
  %``Reconciling the X(3872) with the near-threshold enhancement in the D0 anti-D*0 final state,''
  Phys.\ Rev.\ D, 2007, \textbf{76}: 034007
  %[arXiv:0704.0605 [hep-ph]].
  %%CITATION = ARXIV:0704.0605;%%
  %81 citations counted in INSPIRE as of 03 Jan 2014


%\cite{Kalashnikova:2009gt}
\bibitem{Kalashnikova:2009gt}
  Kalashnikova Y S, Nefediev A V.
  %``Nature of X(3872) from data,''
  Phys.\ Rev.\ D, 2009, \textbf{80}: 074004
  %[arXiv:0907.4901 [hep-ph]].
  %%CITATION = ARXIV:0907.4901;%%
  %32 citations counted in INSPIRE as of 03 Jan 2014


%\cite{Cleven:2013sq}
\bibitem{Cleven:2013sq}
  Cleven M, Wang Q, Guo F K et al.
  %``Confirming the molecular nature of the $Z_b(10610)$ and the $Z_b(10650)$,''
  Phys.\ Rev.\ D, 2013, \textbf{87}: 074006
  %[arXiv:1301.6461 [hep-ph]].


%\cite{Wang:2013cya}
\bibitem{Wang:2013cya}
  Wang Q, Hanhart C, Zhao Q.
  %``Decoding the riddle of Y(4260) and $Z_c(3900)$,''
  Phys.\ Rev.\ Lett.,\ 2013, \textbf{111}: 132003
  %[arXiv:1303.6355 [hep-ph]].


%\cite{Guo:2013zbw}
\bibitem{Guo:2013zbw}
  Guo F K, Hanhart C, Mei{\ss}ner U G et al.
  %``Production of the X(3872) in charmonia radiative decays,''
  Phys.\ Lett.\ B, 2013, \textbf{725}: 127
  %[arXiv:1306.3096 [hep-ph]].


%\cite{Fleming:2007rp}
\bibitem{XEFT}
  Fleming S, Kusunoki M, Mehen T et al.
  %``Pion interactions in the $X(3872)$,''
  Phys.\ Rev.\ D, 2007, \textbf{76}: 034006
  %[hep-ph/0703168].
  %%CITATION = HEP-PH/0703168;%%
  %58 citations counted in INSPIRE as of 03 Jan 2014


\bibitem{Bfactory1}
  Aushev T et al. (Belle Collaboration).
  %``Study of the B ---> X(3872)(D*0 anti-D0) K decay,''
  Phys.\ Rev.\ D, 2010, \textbf{81}: 031103
  %[arXiv:0810.0358 [hep-ex]].


%\cite{Aubert:2007rva}
\bibitem{Bfactory2}
  Aubert B et al. (BaBar Collaboration).
  %``Study of Resonances in Exclusive B Decays to anti-D(*) D(*) K,''
  Phys.\ Rev.\ D, 2008, \textbf{77}: 011102
  %[arXiv:0708.1565 [hep-ex]].
  %%CITATION = ARXIV:0708.1565;%%
  %145 citations counted in INSPIRE as of 03 Jan 2014


%\cite{Beringer:1900zz}
\bibitem{PDG}
  Beringer J et al. (Particle Data Group Collaboration).
  %``Review of Particle Physics (RPP),''
  Phys.\ Rev.\ D, 2012, \textbf{86}: 010001
  %%CITATION = PHRVA,D86,010001;%%
  %2964 citations counted in INSPIRE as of 03 Jan 2014


%\cite{Aubert:2008gu}
\bibitem{Aubert:2008gu}
  Aubert B et al. (BaBar Collaboration).
  %``A Study of $B \to X(3872) K$, with $X_{3872} \to J/\Psi \pi^{+} \pi^{-}$,''
  Phys.\ Rev.\ D, 2008, \textbf{77}: 111101
  %[arXiv:0803.2838 [hep-ex]].
  %%CITATION = ARXIV:0803.2838;%%
  %90 citations counted in INSPIRE as of 03 Jan 2014


%\cite{Adachi:2008te}
\bibitem{Adachi:2008te}
  Adachi I et al. (Belle Collaboration).
  %``Study of $X(3872) $ in $B$ meson decays,''
  arXiv:hep-ex/0809.1224
  %%CITATION = ARXIV:0809.1224;%%
  %60 citations counted in INSPIRE as of 03 Jan 2014


\bibitem{Meng:2013gga}
  Meng C, Han H, Chao K T.
  %``X(3872) and its production at hadron colliders,''
  arXiv:hep-ph/1304.6710
  %%CITATION = ARXIV:1304.6710;%%
  %2 citations counted in INSPIRE as of 27 Oct 2013











\end{thebibliography}
\end{document}